Author:      A. Raymond Penner

Address:     Department of Physics,
             Vancouver Island University,
             900 Fifth Street,
             Nanaimo, BC, Canada,
             V9R 5S5

Email:       raymond.penner@viu.ca

Tel:         250 753-3245 ext: 2336

Fax:         250 740-6482

Gravitational anti-screening and binary galaxies

Abstract

Previously, in Penner (2016a, 2016b), a theory of gravitational anti-screening was shown to lead naturally to the Baryonic Tully-Fisher Relationship. In addition, it was shown to agree with the observed rotational curve of the Galaxy, the observed features in the rotational curves of other spiral galaxies, with observations of the Coma cluster, and with a geometrically flat universe. In this paper the theory will now be applied to binary galaxies. It is shown that there is a relationship between the line-of-sight velocity difference of the pair and the individual rotational velocities of the galaxies. The resulting probability function for β, defined as the ratio of the line-of-sight velocity difference to the rotational velocity of the larger galaxy of the pair, is in excellent agreement with the observations taken by multiple researchers for the case of the binaries being on radial orbits.

## 1. Introduction

Previously the author presented (Penner 2016a) a theory of gravitational anti-screening as an alternative to dark matter. The theory was shown to lead naturally to the Baryonic Tully-Fisher Relationship (BTFR), to be in good agreement with the rotation curve of the Galaxy, to agree with the observed features seen in the rotational curves of other spiral galaxies, and to agree with the velocity dispersions and shear values for the Coma cluster. In Penner (2016b), the theory was extended to superclusters where it was shown to be also in agreement with an overall density parameter equal to 1 and to be compatible with an accelerating expansion without the need of dark energy. The theory of gravitational anti-screening will now be extended to the analysis of binary galaxies.

The analysis of binary galaxies would seem to be an excellent way of testing the current theory of dark matter or alternatives to it such as gravitational anti-screening. However, it does suffer from two major complications. First, observations only provide $v_{los}$, the line-of-sight velocity difference between the two galaxies, and $s_p$, their projected separations. Combined with this the uncertainties and resulting scatter in the data sets are typically large. The second complication is being able to select a representative set of binary galaxies. The primary criteria for selecting binary galaxies in the various studies is their isolation, determined by their angular separation as compared with their angular separation from other galaxies. It is from their angular separation that $s_p$ is calculated. A low cutoff value will eliminate most of the chance projections but limiting the separation of the pairs can bias the results towards certain types of orbits. The second criteria for selecting binary galaxies is setting a cutoff value for $v_{los}$. This cutoff value is often set ad hoc and again there is a tradeoff between eliminating chance projections and biasing the results.

Analysis of the different selected sets of binary galaxies then consists of looking for relationships among the observables, i.e. $v_{los}$, $s_p$, and luminosity, and comparing these results to various models. The models used in the analysis are all based on the concept of dark matter halos surrounding each individual galaxy. Typically, it is taken that the dark matter halos do not interact and the galaxies can be treated as point masses executing Keplerian orbits. The analysis then tries to answer the question; "does such a model agree with the observations and if so what combination of orbital eccentricities and galaxy masses best fit?". Analysis becomes quite complex and statistical in nature and the correlations between the observables and agreements between the various models and observations are generally marginal if there at all. A summary of the major conclusions reached by different researchers based on six different sets of binary galaxies is as follows;

i)      Turner's (1976a, 1976b) selected binary galaxies have projected separations $s_p \leq 200$ kpc ($H_o = 70$ km s$^{-1}$ Mpc$^{-1}$ for this and the other sets). The best fit model was one with moderate-eccentric Keplerian orbits and average total mass to light ratios for the binary galaxies of ~65 $M_\odot/L_\odot$ to ~100 $M_\odot/L_\odot$ (Turner & Ostriker 1977), much greater than galaxy estimates based on their rotational curves. However, it was also concluded that moderate-eccentricity orbits of galaxies with massive halos are too complex to model reliably as it is nearly impossible to fully untangle the effects of orbit eccentricity and halo size.

White (1981) and White et al (1983) used Turners' same set of binary galaxies, albeit with more precise velocities, to determine possible power law relationships between $v_{los}$, $s_p$, and total pair luminosities. There was found to be no relationship between $v_{los}$ and the projected separations. It was concluded that Keplerian orbits do not fit the observations. The interaction between the pairs corresponds best to a logarithmic interaction potential with the effective mass-to-light ratio increasing linearly with the pair's separation.

The difference between the conclusions reached by Turner and White et al, although using the same set of binaries, directly highlight the difficulty in drawing conclusions from a statistical analysis of binary galaxy observations.

ii)     Peterson's (1979a, 1979b) selected set of binary galaxies have $s_p \leq 600$ kpc. The line-of-sight velocity difference was found to be independent of the projected separation out to 100 kpc, but then fell off with distance. Peterson just considered a point-mass circular orbit model and determined an estimate of the average total mass and mass to light ratio for the pairs for such orbits. The best estimate was that $M/L \sim 50$ $M_\odot/L_\odot$. Again, it is much greater than galaxy estimates based on their rotational curves.

iii)    Van Moorsel's (1982, 1983a-c, 1987) selected set of binary galaxies have $s_p \leq 250$ kpc. Both a point mass model of the galaxies and a model where the matter distribution of the two galaxies, based on their rotation velocities, was extrapolated until the two distributions touched were considered. The orbits for both these models were taken to be circular. It was found that the point mass model differed greatly from the observations while the model with the extrapolated mass distribution was found to be in reasonable agreement. This model with the extrapolated mass distributions was shown to be equivalent to a logarithmic interaction potential.

iv)     Schweizer's (1987a, 1987b) selected set of binary galaxies have $s_p \leq 150$ kpc. It was found that beyond 100 kpc the $v_{los}$ values were significantly lower than at the smaller separations. Models with moderately eccentric Keplerian orbits were found to marginally represent the data better than models with nearly circular orbits. Estimated masses were approximately 5 times those estimated from galactic rotation curves with the mass increasing approximately linearly with radius although not as far as their mean separation.

v)      Chengalur et al (1993) analyzed binary galaxies with projected separations of up to 1.0 Mpc. The average $v_{los}$ was significantly lower as compared to other studies which typically considered much smaller projected separations. This result was taken to be consistent with these wide pairs being on almost radial orbits with very large maximum separations.

vi)     Honma (1999) does not directly provide the projected separations for the data set, but it is estimated that $s_p \leq 400$ kpc. Both circular and isotropic orbits were considered in the analysis. It was found that M/L was almost independent of separation for $s_p > 100$ kpc. It was concluded that the typical halo size of spiral galaxies is therefore on the order of 100 kpc.

As is seen with this sample of studies, there is no conclusive answers regarding the dynamics of binary galaxies. The problem is partially related with the complications mentioned above but it is also with the theoretical models being used. The researchers are trying to fit the observations to models where galaxies have dark matter halos of set size executing Keplerian orbits. In the author's opinion, researchers are trying to adapt a flawed theory to the observations. As will be demonstrated in this manuscript, the theory of gravitational anti-screening will lead to results that are consistent with all six sets of binary galaxies and will provide a definitive answer to the dynamics of binary galaxies.

## 2. Theory

### 2.1 Gravitational anti-screening

The initial motivation behind the theory of gravitational anti-screening was to explain the Baryonic Tully-Fisher Relationship (McGaugh et al 2000, McGaugh 2012). The BTFR is the empirical relationship that exists between $M_B$, the baryonic mass of a galaxy, and $v_r$, the galaxy's constant outer rotational velocity. The BTFR as given by McGaugh (2012) is

$$M_B = A \, v_r^4 \tag{1a}$$

with

$$A = (47 \pm 6) \, M_\odot \, km^{-4} \, s^4. \tag{1b}$$

The BTFR can also be expressed as the following relationship between M, the observed mass of the galaxy, its baryonic mass, and the observation distance r, by substituting $GM/r$ for $v_r^2$ in (1a);

$$M = (G^2A)^{-1/2} r \, M_B{}^{1/2}. \tag{2}$$

In the model of gravitational anti-screening, presented in Penner (2016a), baryonic masses are surrounded by a sea of virtual mass dipoles. This is analogous to quantum electrodynamic theory where charges are surrounded by a sea of virtual electric dipoles. In QED the virtual electric dipoles result in a screening effect that leads to the value of the observed charge of a particle being less than its actual bare charge. In the gravitational case the virtual mass dipoles result in an anti-screening effect and in the case of galaxies leads to the observed mass of the galaxy being greater than its baryonic mass.

In Penner (2016a) the following model of the dependence that $\mathbf{P_G}$, the virtual mass dipole density surrounding a baryonic mass, has on $\mathbf{g}$, the total gravitational field, is used;

$$\mathbf{P_G} = \frac{1}{4\pi G} g_o \, (1 - e^{-g/g_o}) \, \hat{\mathbf{g}} \, , \tag{3}$$

where $g_o$ is a model parameter that is determined empirically. The resulting induced mass density of the vacuum, $\rho_v$, surrounding the given baryonic mass is then given by;

$$\rho_v = -\boldsymbol{\nabla}\cdot\mathbf{P_G} \, . \tag{4}$$

This induced mass density contributes to the gravitational field as with any other mass component and the gravitational field due to this induced mass density is thereby determined from

$$\mathbf{g_v} = - \, G \int_V \frac{\rho_v \hat{\mathbf{r}}}{r^2} \, dV \, . \tag{5}$$

As shown in Penner (2016a), for individual galaxies this model leads to the following far field relationship between the galaxy's total observed mass, its baryonic mass, and the observation distance;

$$M = \left(\frac{2g_o}{G}\right)^{1/2} r \, M_B{}^{1/2}. \tag{6}$$

Comparing (6) to (2) it is seen that the theory of gravitational anti-screening leads to the BTFR. The value of the parameter $g_o$ was then obtained by equating the coefficients of (6) and (2);

$$g_o = \frac{1}{2GA} = (8.0 \pm 1.0) \times 10^{-11} \text{ m s}^{-2}. \tag{7}$$

As mentioned in the introduction, in addition to leading to the BTFR, the theory of gravitational anti-screening was shown to be in good agreement with the rotation curve of the Galaxy, observed features in the rotational curves of other spiral galaxies, the velocity dispersions and shear values for the Coma cluster (Penner 2016a). In Penner (2016b) it was further shown to agree with the overall density parameter being equal to 1. The same technique that was used to determine the rotation curves of galaxies can be applied to binary galaxies. The initial estimate of the total gravitational field is taken to be solely that due to the baryonic masses of the two galaxies. The value of $\mathbf{P_G}$ and the resulting induced mass density distribution of the vacuum is then determined using (3) and (4). For the next estimate, the total gravitational field is then taken to be equal to the sum of the fields due to the baryonic masses and the vacuum contribution as given by (5). This iterative process is repeated until the resulting values of the total gravitational field obtained after a given iteration vary by less than a set amount from the previous iteration.

Figures 1 and 2 show the resulting induced mass distributions around a pair of galaxies with baryonic masses $M_{B1} \gg M_{B2}$ and $M_{B2} = M_{B1}$ respectively. In both figures $M_{B1} = 60$ x $10^9$ $M_\odot$ and the two galaxies are separated by s = 100 kpc. Also in both figures the induced mass distribution is only shown out to where it has dropped to a density of $10^{-24}$ kg m$^{-3}$. The distribution associated with a particular galaxy or with a given pair does not terminate as with dark matter halos but will extend to as far as to where the gravitational field from the given baryonic mass is dominant. As the pair's separation changes the distribution changes. It needs to be stressed that Figure 2 is not showing "halos" interacting as with dark matter. These distributions that are induced from the vacuum are solely dependent on the local gravitational fields as per (3) and (4).

In the following analysis the two extreme relationships between the galactic baryonic masses will be considered, i.e. $M_{B1} \gg M_{B2}$ and $M_{B2} = M_{B1}$. Results obtained from these two extreme cases would be expected to determine the bounds of the more general case. It will also be taken in the analysis that the separation of the two galaxies is great enough such that they can be treated as point masses.

## 2.2 Orbits for $M_{B1} \gg M_{B2}$

In the case where $M_{B1} \gg M_{B2}$, the effect that the smaller galaxy has on the induced mass distribution surrounding the larger galaxy is negligible. This can be seen with Figure 1. The gravitational field at the smaller galaxy due to the larger galaxy will then be given by

$$g_{21} = \frac{GM_1}{s^2} \tag{8}$$

where s is the separation of the two galaxies and $M_1$ is the total mass within a distance s of the larger galaxy. Substituting from (6), (8) becomes

$$g_{21} = \frac{(2g_o G)^{1/2} M_{B1}^{1/2}}{s} \tag{9}$$

and the acceleration of the smaller galaxy is therefore inversely proportional to the separation. The gravitational field at the smaller galaxy for this case can also be expressed as

$$g_{21} = \frac{v_{r1}^2}{s} \tag{10}$$

where $v_{r1}$ is the rotational velocity of the larger galaxy.

A central force which is inversely proportional to the separation leads to a logarithmic interaction potential. A complication for the analysis for such a force is that it does not, in general, lead to closed orbits. Therefore, only the two special cases that do lead to closed orbits will be considered, namely circular orbits and purely radial orbits.

### 2.2.1 Circular orbits

For $M_{B1} >> M_{B2}$ the centre of mass of the system can be taken to be located at the centre of $M_{B1}$. If the smaller galaxy is in circular orbit around the larger galaxy, the velocities of the galactic centres with respect to the centre of mass will then be given by

$$|\mathbf{v_1}| = 0 \tag{11a}$$

$$|\mathbf{v_2}| = v_{r1} \tag{11b}$$

and $\mathbf{v_D}$, the velocity difference of their centres, by

$$|\mathbf{v_D}| = |\mathbf{v_2} - \mathbf{v_1}| = v_{r1}. \tag{12}$$

The smaller galaxy is basically equivalent to a star orbiting the larger galaxy.

As discussed in the introduction, observations of binary galaxies provide $v_{los}$, the the line-of-sight velocity difference between their galactic centres. For an angle $\theta$ between $\mathbf{v_D}$ and the line-of-sight from the Earth,

$$v_{los} = v_D \cos \theta. \tag{13}$$

If there is no preferred direction for $\mathbf{v_D}$, so that in a representative set of binary galaxies the $\mathbf{v_D}$'s are equally distributed in all directions, the probability function for $\theta$ will be given by

$$P(\theta) = \sin \theta. \tag{14}$$

From (13) and (14) the probability function for the line-of-sight velocity will then be given by

$$P(v_{los}|v_D) = P(\theta) \left( \frac{-dV_{los}}{d\theta} \right)^{-1} \tag{15a}$$

$$= \frac{1}{v_D} \ , \qquad 0 \leq v_{los} \leq v_D \tag{15b}$$

or equivalently, given (12),

$$P(v_{los}|v_{r1}) = \frac{1}{v_{r1}} \ . \qquad 0 \leq v_{los} \leq v_{r1} \tag{16}$$

Defining

$$\beta = v_{los}/v_{r1}, \tag{17}$$

the probability function for $\beta$ will be given by

$$P(\beta) = P(v_{los}|v_{r1}) \frac{dV_{los}}{d\beta} \ . \tag{18}$$

For circular orbits it follows from (16-18) then that

$$P(\beta) = 1 \qquad 0 \le \beta \le 1. \tag{19}$$

with the average value of $\beta$ for these orbits given by

$$\bar{\beta} = \int_0^1 \beta \, P(\beta) d\beta = 0.50 \,. \tag{20}$$

### 2.2.2 Radial orbits

In the case of radial orbits the smaller galaxy will oscillate along the line joining the two galaxies. The centre of mass of the system will again be at the centre of the larger galaxy and the velocity difference between the pair will then equal the radial velocity of the smaller galaxy. Integrating (10), $v_D$ is determined to be

$$v_D = 2^{1/2} \, v_{r1} \left(\ln\left(\frac{s_o}{s}\right)\right)^{1/2} \tag{21}$$

where $s_o$ is the initial or maximum separation of the two galaxies. Figure 3 shows the resulting dependence that $v_D$ has on s. The separation of the two galaxies is in turn determined by inverting (21), to be

$$s = s_o \, e^{-\left(\frac{v_D^2}{2v_{r1}^2}\right)}. \tag{22}$$

The probability function for the smaller galaxy being found at a distance s ($s_o \ge s \ge 0$) from the larger galaxy is given by

$$P(s) = \frac{4}{T} \frac{1}{v_D} \tag{23}$$

where T is the period of oscillation and T/4 is the time it takes s to go from $s_o$ to 0. The resulting probability function for $v_D$ is by (22) and (23) then given by

$$P(v_D|v_{c1}) = P(s)\left(-\frac{ds}{dv_D}\right) \tag{24a}$$

$$= \frac{4s_o}{Tv_{r1}^2} \, e^{-\left(\frac{v_D^2}{2v_{r1}^2}\right)}. \tag{24b}$$

From the normalization condition for $P(v_D|v_{c1})$ the following expression for T is derived

$$T = (8\pi)^{1/2} \frac{s_o}{v_{r1}} \,. \tag{25}$$

Substituting (25) into (24b) then results in

$$P(v_D|v_{r1}) = \left(\frac{2}{\pi}\right)^{1/2} \frac{1}{v_{r1}} \, e^{-\left(\frac{v_D^2}{2v_{r1}^2}\right)}. \tag{26}$$

By (15b) and (26) it then follows that

$$P(v_{los}|v_{r1}) = \int_{v_{los}}^{\infty} P(v_{los}|v_D) \, P(v_D|v_{r1}) dv_D \tag{27a}$$

$$= \left(\frac{2}{\pi}\right)^{1/2} \frac{1}{v_{r1}} \int_{v_{los}}^{\infty} \frac{1}{v_D} e^{-\left(\frac{v_D^2}{2v_{r1}^2}\right)} dv_D \tag{27b}$$

$$= -\left(\frac{1}{2\pi}\right)^{1/2} \frac{1}{v_{r1}} \left[\text{Ei}\left(-\frac{v_{los}^2}{2v_{r1}^2}\right)\right] \tag{27c}$$

where Ei is the exponential integral. From (18) the probability function for $\beta$ for $M_{B2} \ll M_{B1}$ and for radial orbits is then given by

$$P(\beta) = -\left(\frac{1}{2\pi}\right)^{1/2} \left[\text{Ei}\left(-\frac{\beta^2}{2}\right)\right] \tag{28}$$

with the average value of $\beta$ for these orbits given by

$$\bar{\beta} = \int_0^1 \beta\, P(\beta) d\beta = \left(\frac{1}{2\pi}\right)^{1/2} \cong 0.40. \tag{29}$$

The probability functions for $\beta$ for $M_{B1} \gg M_{B2}$ and for both circular orbits, (19), and radial orbits, (28), are shown on Figure 4.

## 2.3 Orbits for $M_2 = M_1$

In Figure 2 it is seen that for $r < s/2$ (50 kpc) each galaxy is surrounded by its own localized induced mass distribution. However, as r increases the distributions morph into a single distribution equivalent to that surrounding a single baryonic mass equal to $M_{B1} + M_{B2}$. In the case shown, where $M_{B2} = M_{B1}$, the localized induced mass distributions around each of the two galaxies extends to the mid-point, which will be the centre of mass for the system. As an approximation, it will be modeled that the induced mass distribution for each galaxy extends undisturbed to this midpoint. In the case of circular orbits, with $M_{B2} = M_{B1}$, each galaxy will execute orbits of radius s/2 about the midpoint. The gravitational field at each galactic centre due to the other galaxy will be given by (8), but in this case $M_1$ will be the total mass only within s/2 of the galaxy. Substituting from (6), (8) then becomes

$$g_{21} = \frac{(2g_0 G)^{1/2} M_{B1}^{1/2}}{2s} \tag{30}$$

and (10) becomes

$$g_{21} = \frac{v_{r1}^2}{2s}. \tag{31}$$

The individual galactic velocities about the centre of mass will be given by

$$|\mathbf{v_1}| = |\mathbf{v_2}| = \frac{v_{r1}}{2} \tag{32a}$$

$$\mathbf{v_2} = -\mathbf{v_1} \tag{32b}$$

and the velocity difference of their centres by

$$|\mathbf{v_D}| = |\mathbf{v_2} - \mathbf{v_1}| = v_{r1}. \tag{33}$$

The velocity difference is the same as given by (12) and the derivation provided in 2.2.1 for circular orbits applies. Therefore for both $M_{B1} \gg M_{B2}$ and $M_{B2} = M_{B1}$ and circular orbits, the probability function for $\beta$ and the value for $\bar{\beta}$ is as given by (19) and (20) where $\beta$ is the ratio of $v_{los}$ to the rotational velocity of the larger galaxy.

In the case of radial orbits, the magnitude of the acceleration of each of the two equal mass galaxies towards each other will be given by (31) and the relative acceleration between the two will therefore be as given by (10). The velocity difference will therefore be as given by (21) and the derivation provided in 2.2.2 for radial orbits therefore applies. Therefore for both $M_{B1} \gg M_{B2}$ and $M_{B2} = M_{B1}$ and radial orbits, the probability function for $\beta$ and the value for $\bar{\beta}$ is as given by (28) and (29).

In the above derivation it was approximated that the localized induced mass distributions of each galaxy remains undisturbed up to the mid-point of the pair. In reality, the induced mass distribution around each galaxy is influenced by the gravitational field of the other. This is clearly seen in Figure 2, which is the case, $M_{B2} = M_{B1}$, where the influence is the greatest. To take the distortion of the induced mass distribution into account, (31) will be modified as such

$$g_{21} = \gamma^2 \frac{v_{r1}^2}{2s} \tag{34}$$

with $\gamma$ being a measure of the effect of the distortion. Carrying $\gamma$ through the derivations then results in, for circular orbits;

$$P(\beta) = \frac{1}{\gamma} \qquad 0 \le \beta \le \gamma \quad \text{with } \bar{\beta} = 0.50 \, \gamma \, . \tag{35}$$

and for radial orbits;

$$P(\beta) = -\left(\frac{1}{2\pi}\right)^{1/2} \frac{1}{\gamma} \left[ \text{Ei}\left(-\frac{\beta^2}{2\gamma^2}\right) \right] \quad \text{with } \bar{\beta} = \gamma \left(\frac{1}{2\pi}\right)^{1/2} \cong 0.40 \, \gamma \, . \tag{36}$$

In order to determine the value of $\gamma$, simulations were run for different values of $M_{B2} = M_{B1}$ and different $s_p$'s. The value of $\gamma$ was found to be relatively constant with values ranging from 1.20 to 1.25. The theoretical value of $\bar{\beta}$ therefore ranges from 0.50 to 0.625 for circular orbits and from 0.40 to 0.50 for radial orbits.

The probability functions for $\beta$ for $M_{B2} = M_{B1}$ and with $\gamma = 1.25$, for both circular and radial orbits, are included on Figure 4. As is seen on the figure there is little difference between the distributions for $M_{B2} \ll M_{B1}$ and $M_{B2} = M_{B1}$. As such the average of the two distributions, determined by setting $\gamma = 1.125$ in (35) and (36), for both circular and radial orbits will represent the theoretical distributions when comparing theory to the results.

## 3. Results

In order to compare the different sets of binary galaxies with the theory, the rotational velocity of the larger galaxy of a given pair is needed in addition to $v_{los}$. Petersen provides H1 profile widths corrected for inclination for 33 galaxy pairs and Chengalur et al provides H1 profile widths corrected for inclination for 32 galaxy pairs. In both these cases the rotational velocity of each galaxy is taken to be one-half of the profile width. In the case of van Moorsel the maximum rotational velocity of each galaxy is directly provided for the 14 galaxy pairs. The values of β for each of these 79 pairs were therefore able to be determined directly. For the Turner/White, Schweizer, and Honma sets neither the H1 profile widths nor the rotational velocities of the galaxies were provided. In these cases, the values for $v_r$ and β were determined indirectly by using the following Tully-Fisher relationship (Sakai et al 2000) for spiral galaxies,

$$v_r = 200 \text{ km s}^{-1} \left( \frac{L}{2.97 \text{ x} 10^{10} L_\odot} \right)^{0.277}$$

where L is the V-band luminosity of the galaxy. In Schweizer's case the V-band luminosity of each galaxy was provided while for both Turner/White and Honma's sets the V-band luminosities were looked up in NED. For these sets only the pairs where the larger galaxy (taken to be the more luminous) was of late type, was identified so it could be looked up in NED, and had a visual luminosity provided for in NED, were used. This resulted in 30 binary galaxies for Schwizer, 53 pairs for Turner/White, and 46 pairs for Honma.

Table 1 lists each of the six sets of binary galaxies along with their resulting $\bar{\beta}$'s. As is seen the values of $\bar{\beta}$ are consistent across the sets except for that belonging to Chengalur et al. This will be discussed later. The value of $\bar{\beta}$ for all 208 binary galaxies of the six sets is 0.512. This value of $\bar{\beta}$ is however very sensitive to outliers, which are most likely due to chance projections. Out of the 208 binary galaxies, two values of β exceeded 3. If these two outliers are removed the value for $\bar{\beta}$ for the remaining 206 pairs drops to 0.488. Overall the values of $\bar{\beta}$ are in excellent agreement with the range of values determined from the theory of gravitational anti-screening.

To determine the nature of the orbits the probability function P(β) was determined for each of the six sets of binary galaxies. On Figure 5 are shown the results for the three sets, Petersen, Chengalur et al, and Honma, which included pairs of wide separation, i.e. $s_p$'s up to 400 kpc and greater. As seen they are in very good agreement with each other. On Figure 6 is the combined P(β) for all three sets along with the theoretical P(β) (i.e. using γ = 1.125) for both circular and radial orbits. The result is very definitive, the theory of gravitational anti-screening is in excellent agreement with these three sets of binary galaxies being in radial orbits. The specific results of Chengalur et al adds further weight to this conclusion as they had considered pairs of much wider separation than the other groups. For radial orbits this would correspond to having pairs with a greater probability of being perpendicular to the line-of-sight and/or having $s/s_o$ closer to 1. Either case leads to lower $v_{los}$'s and lower β's. As Figure 5 shows Chengalur et al's pairs have a higher number of low β's than the other sets and a correspondingly lower $\bar{\beta}$, as was seen in Table 1.

The results for the Turner/White, Schweizer, and van Moorsel binary data sets, where only relatively closely spaced pairs were considered, i.e. $s_p \leq 250$ kpc, are shown on Figure 7. The three are in very good agreement with each other although for the 14 binary galaxies of van Moorsel's set the scatter is quite large. Especially interesting is the excellent agreement between Turner/White and Schweizer where the researchers using the dark matter halo model reached different conclusions with their analysis. On Figure 8 is the combined P(β) for all three of these sets along with the theoretical P(β) for both circular and radial orbits. The result is not as clear as with Figure 6 as there are significantly fewer pairs with β < 0.2. However, the distribution is what would be expected if the binary galaxies are on radial orbits and the low cutoff for $s_p$ eliminates many of those that are nearly perpendicular to the line-of-sight i.e. many of those which will have a small $v_{los}$ and therefore a small β. These three sets, with their relatively small cutoffs for $s_p$, will not then be true representative sets of binary galaxies. For completeness, Figure 9 shows the combined probability function for all six sets of binary galaxies along with the theoretical P(β)'s.

The conclusion that binary galaxies are in radial orbits is also consistent with the previous application of the theory of gravitational anti-screening to the Coma cluster (Penner 2016a, 2014). There it was found by considering the velocity dispersion of galaxies within the cluster that the theory of gravitational anti-screening fitted with the galaxies also being in radial orbits.

Further observational evidence for binary galaxies being in radial orbits comes from the one pair of galaxies where **v_D** is known. Van der Marel et al (2012) determined the velocity vector of M31 with respect to the Galaxy. They found a radial velocity of M31 with respect to the Galaxy of -(109.3 $\pm$ 4.4) km s$^{-1}$, and a tangential velocity of 17.0 km s$^{-1}$. They concluded that the velocity of M31 is statistically consistent with a radial (head-on collision) orbit towards the Galaxy. With a value for $v_{r1} \cong 250$ km s$^{-1}$ (corresponding to the rotational velocity of the larger M31) resulting in $v_D/v_{r1} = 0.44$, M31 and the Galaxy would appear to be a very typical binary. Integrating (21), and using s = 780 kpc for the current separation of the two, it is determined that M31 and the Galaxy will collide, i.e. s → 0, in another 2.73 x 10$^9$ years. Of course a proper treatment of M31 and Galaxy would need to take into account the gravitational fields of other members of the local group.

## 4. Conclusion

Applying the theory of gravitational anti-screening to binary galaxies leads to a relationship between the line-of-sight velocity difference and the rotational velocities of the pair. The resulting probability distribution function for β, defined as the ratio of the line-of-sight velocity difference to the rotational velocity of the larger galaxy of the pair, is in excellent agreement with the observations taken by multiple researchers for the case of the binaries being on radial orbits. Indeed, if the different research groups had all used the gravitational anti-screening model in their analysis they would most likely have had the same conclusions. Unfortunately, they used models based on fixed size dark matter halos resulting in marginal fits at best and a wide range of conclusions.

It is important to point out that the data sets used in the manuscript were not selected for agreement with the theory. These are a range of binary data sets that were found in the literature. It is hoped that independent researchers will produce other binary data sets, specifically ones that include the individual galactic rotation velocities. The author is confident that the more observations included it will be found that the better the fit is with the theory.

The theory of gravitational anti-screening has now been shown to be a better match than the standard dark matter halo model to observations of the rotational curves of galaxies, the dynamics of the Coma cluster, and the dynamics of binary galaxies. Surprisingly, after many decades the standard dark matter model still relies on free parameters such as halo distribution and size to fit as best it can to individual observations. There are no free parameters with the theory of gravitational anti-screening. The same program is used to generate rotational curves for galaxies, the dynamics of a cluster, the density parameter for the universe, or the dynamics of binary galaxies. The only input to the program is the baryonic mass distribution. Unfortunately, the dark matter model, with its great flexibility, would seem to be unfalsifiable.

The theory of gravitational anti-screening is somewhat similar to the theory of dark matter in that the induced dipole contribution provides an equivalent contribution to the gravitational field that would be found from a distribution of dark matter particles. However there are significant differences. First, the gravitational anti-screening theory requires baryonic masses as the source. Without baryonic masses there would be no contribution from the vacuum. Second there is no specific limit to the extent of the distribution surrounding a given baryonic mass. As shown here for binary galaxies the distribution will just morph into one surrounding the pair of galaxies.

The theory of gravitational anti-screening is also somewhat similar to MOND. MOND also leads to the BTFR and the value of $a_o$, the parameter used in MOND, is similar to the value of $g_o$, the parameter used in the theory of gravitational anti-screening. However, this last point is just a result of the value of both parameters falling out from the observational BTFR. More significant is that in MOND the interaction potential will also be logarithmic in nature. As such it would also be expected that MOND would lead to a similar relationships for $\beta$ as derived in this manuscript and would be expected to lead to the same general conclusion as found with the theory of gravitational anti-screening, namely that binary galaxies are dominantly in radial orbits. However a significant difference between the theory of gravitational anti-screening and MOND is in the application to cosmology. As demonstrated in Penner (2016b), the theory of gravitational anti-screening when applied to superclusters leads to an overall density parameter of 1 and does not require dark energy to explain the acceleration of the universal expansion.

The major problem for the theory of gravitational anti-screening is that no specific model for the virtual mass dipoles is presented. Various models for mass dipoles that have been presented over the years are summarized in Penner (2016a). One issue is that mass dipoles, where one of the particles has a negative mass, would seem to violate General Relativity. However, as discussed in Penner (2016a), a solution to Einstein's GR equations which allows for mass dipoles was found (Bondi 1957). At this point I have not incorporated Bondi's solution into the theory of gravitational anti-screening. At this time further work will involve applying the theory to other baryonic mass distributions including the solar system. It is hoped that applying the theory to the solar system will lead to predictions that can be verified.

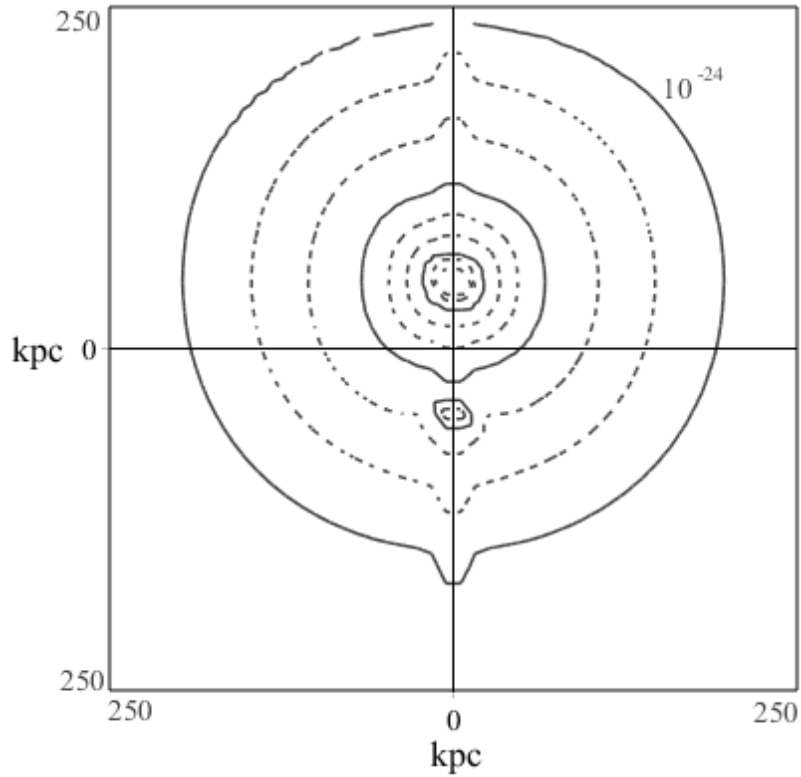

**Figure 1:** The induced mass distribution out to a mass density of $10^{-24}$ kg m$^{-3}$ around a pair of galaxies with $M_{B1} = 60 \times 10^9$ M$_\odot$ , $M_{B2} = 60 \times 10^6$ M$_\odot$ and with the galaxies separated by s = 100 kpc. The induced mass density ranges from $10^{-24}$ kg m$^{-3}$ to $10^{-22}$ kg m$^{-3}$ as one moves from the outer solid line to the inner solid line. (The secondary features along the vertical axis are not real but are a result of computer limitations.)

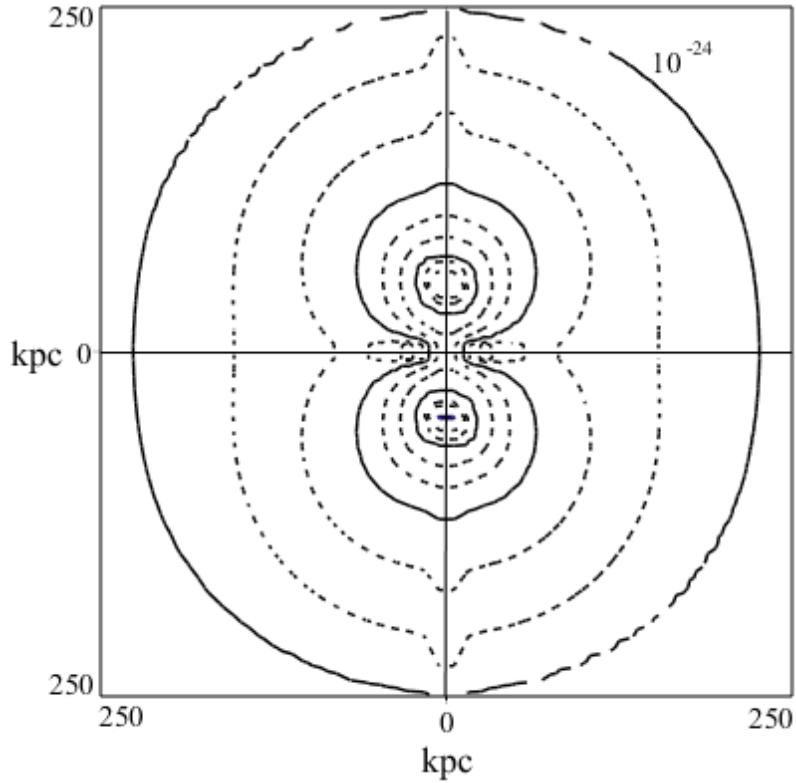

**Figure 2:** The induced mass distribution out to a mass density of $10^{-24}$ kg m$^{-3}$ around a pair of galaxies with $M_{B1} = M_{B2} = 60 \times 10^9$ M$_\odot$ and with the galaxies separated by s = 100 kpc. The induced mass density ranges from $10^{-24}$ kg m$^{-3}$ to $10^{-22}$ kg m$^{-3}$ as one moves from the outer solid line to the inner solid lines. (The secondary features along the vertical axis are not real but are a result of computer limitations.)

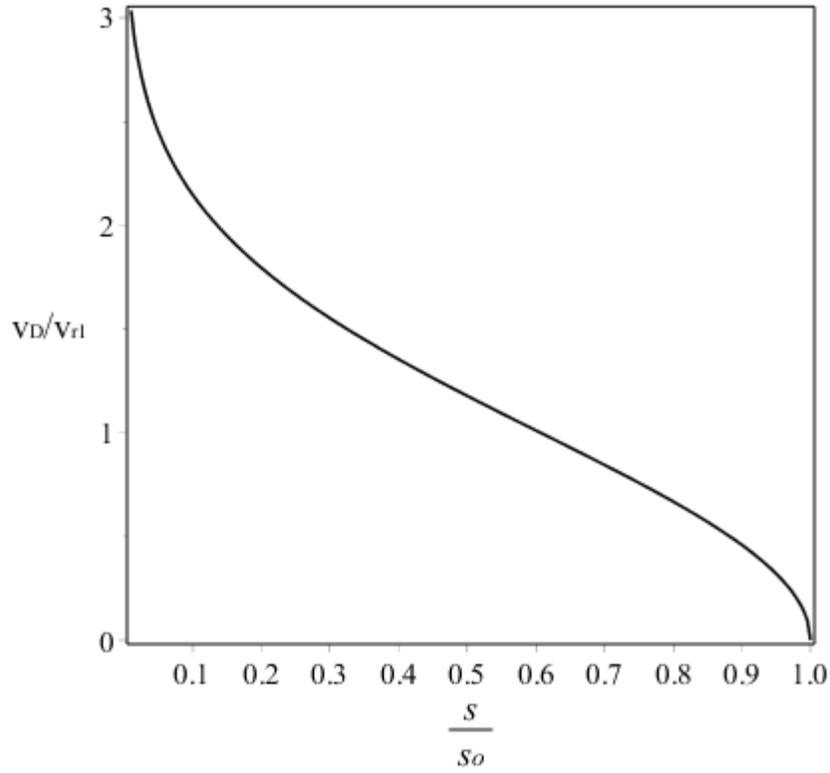

**Figure 3**: The dependence that $v_D$, the magnitude of the velocity difference between binary galaxies, has on s, the separation of the pair, in the case of radial orbits where $v_{rl}$ is the rotational velocity of the larger galaxy and $s_o$ is the initial separation.

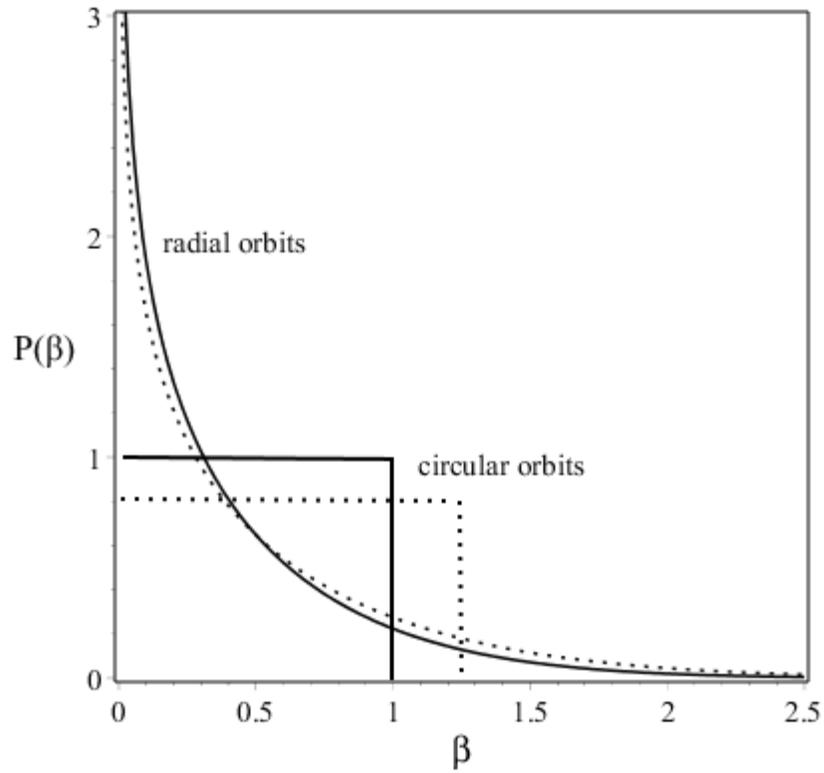

**Figure 4**: The probability function for β = $v_{los}/v_{r1}$ for both circular orbits and radial orbits in the cases of: $M_{B2} << M_{B1}$ ( — ) and $M_{B2} = M_{B1}$ ( ⋯ ).

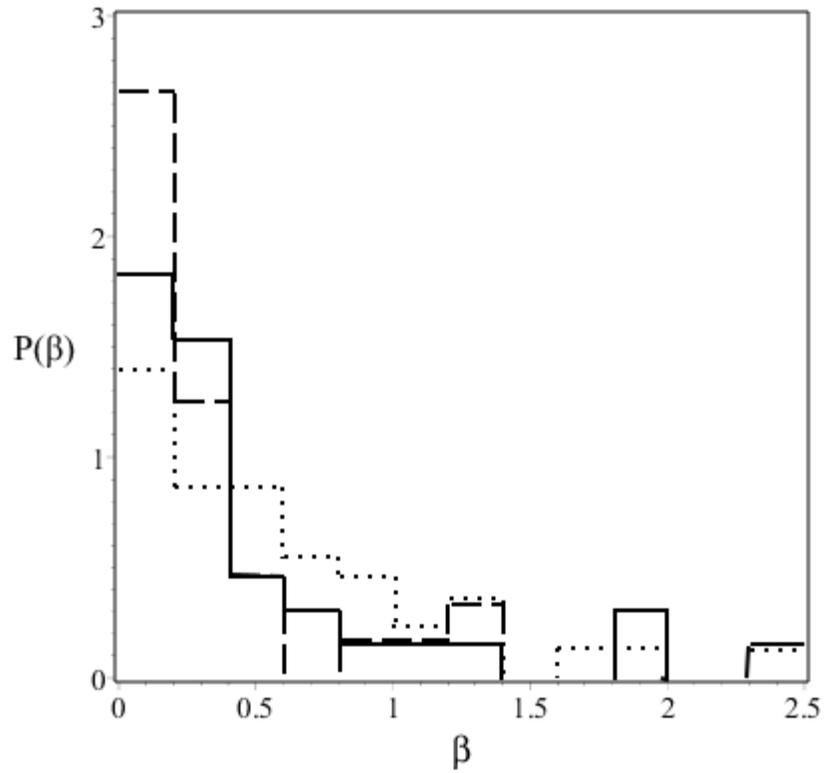

**Figure 5**: The probability function for β = $v_{los}/v_{r1}$ for the binary data sets of Petersen ( — ), Chengalur et al ( --- ), and Honma ( ⋯ ). Values for β > 2 are placed in the last bin.

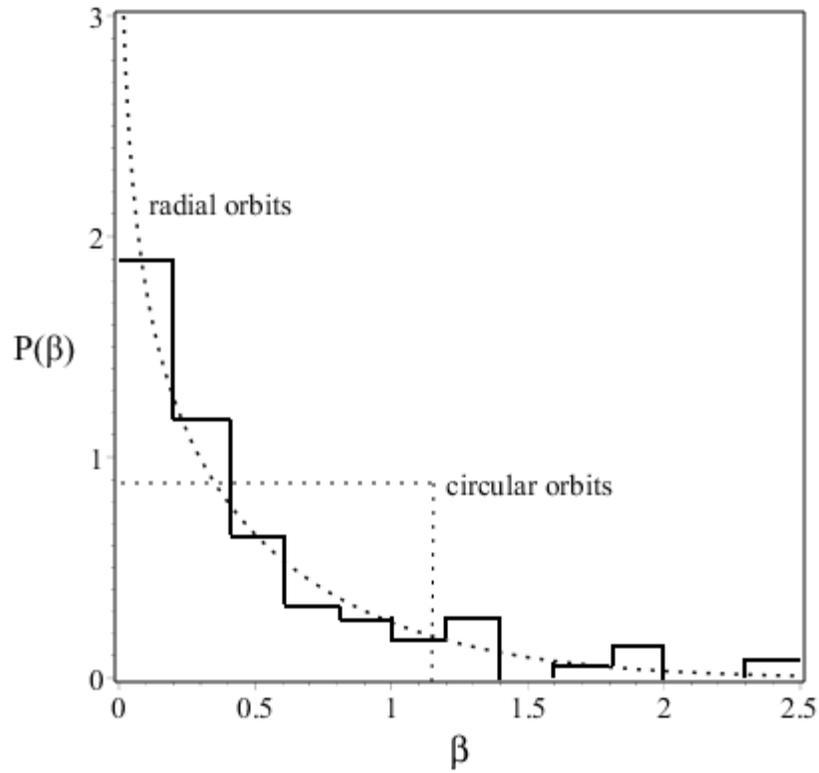

**Figure 6**: The combined probability function for $\beta = v_{los}/v_{r1}$ for the binary data sets of Petersen, Chengalur et al, and Honma along with the theoretical probability functions for both radial and circular orbits. Values for $\beta > 2$ are placed in the last bin.

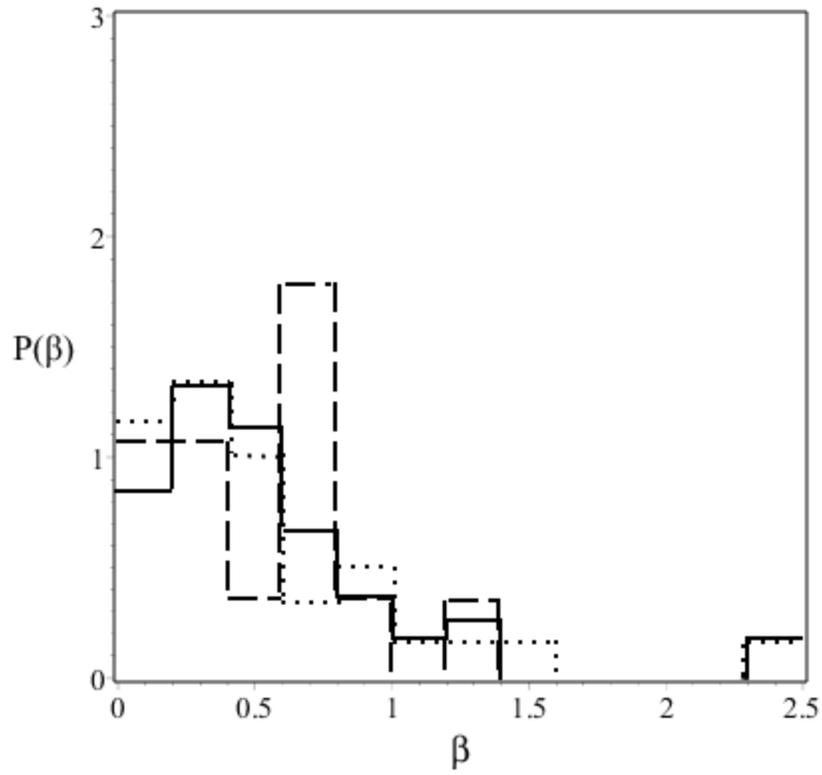

**Figure 7**: The probability function for β = $v_{los}/v_{rl}$ for the binary data sets of Turner/White ( — ), van Moorsel ( --- ), and Schweizer ( ⋯ ). Values for β > 2 are placed in the last bin.

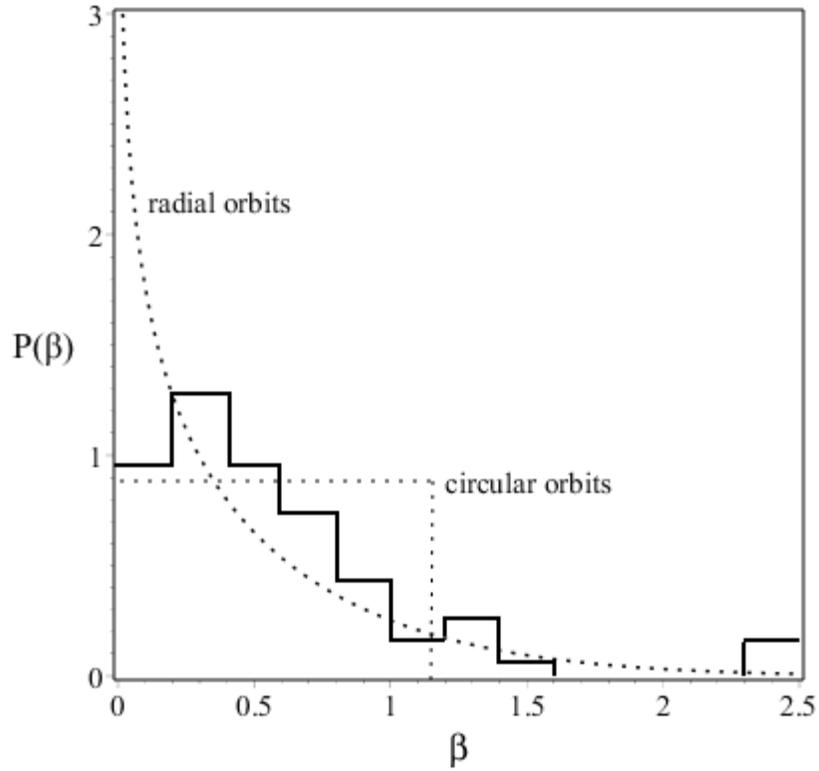

**Figure 8**: The combined probability function for $\beta = v_{los}/v_{r1}$ for the binary data sets of Turner/White, van Moorsel, and Schweizer along with the theoretical probability functions for both radial and circular orbits. Values for $\beta > 2$ are placed in the last bin.

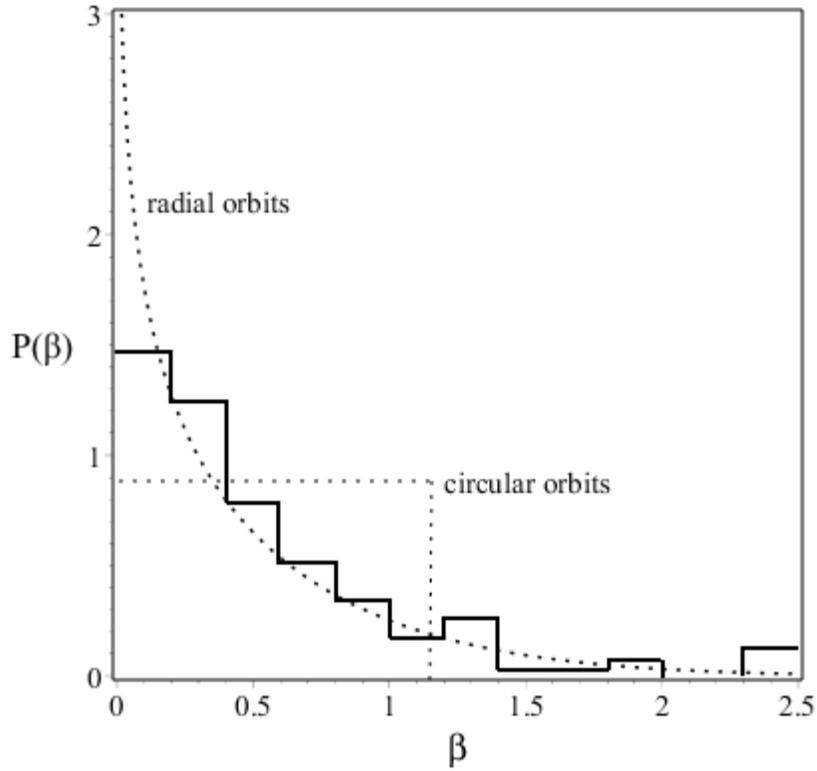

**Figure 9**: The combined probability function for $\beta = v_{los}/v_{r1}$ for the six binary data sets along with the theoretical probability functions for both radial and circular orbits. Values for $\beta > 2$ are placed in the last bin.